# MixiSpeech: Progressive State Distillation for One-Step Vietnamese TTS

Tran Quoc Bao Le[1][0009-0001-9024-5858], Binh Mai[1][0009-0009-9957-8702], Hung Dinh[1][0009-0003-6278-1092], Lam Nguyen[1][0009-0005-3550-9864], and Cong Tran[1][0000-0001-9467-4978]⋆

Posts and Telecommunications Institute of Technology, Hanoi, Vietnam
{k100iltqbao, binhmai2205, dinhhung15082004, hailam04.work}@gmail.com, congtt@ptit.edu.vn

**Abstract.** Conditional-flow text-to-speech (TTS) models can achieve high synthesis quality with relatively few sampling steps, but collapsing the remaining trajectory to a single update can still cause substantial quality degradation. We propose MixiSpeech, a progressive state-distillation framework that converts a multi-step conditional-flow acoustic model into a one-step generator while preserving its architecture and speaker conditioning. Rather than training the one-step model directly, MixiSpeech progressively reduces the global sampling budget through a $10 \rightarrow 4 \rightarrow 2 \rightarrow 1$ curriculum, with each student trained to match local teacher-integration endpoints using a masked state-matching objective. We instantiate the framework on a 710-speaker Vietnamese Matcha-TTS model trained from scratch on PhoAudiobook. On 100 shared test text–speaker pairs, direct one-step inference substantially degrades synthesis quality, obtaining 1.6846 UTMOS22-Strong and 6.21% WER compared with 2.2923 and 4.65% for the 10-step teacher. In contrast, the distilled one-step model achieves 2.2678 UTMOS22-Strong, 4.63% WER, and 2.83% CER, recovering most of the quality lost at the one-step operating point while requiring only a single acoustic vector-field evaluation. End-to-end inference time decreases from 0.0662 s for the 10-step teacher to 0.0312 s for MixiSpeech, corresponding to a 2.12× measured speedup. These results show that progressive state distillation can make one-step conditional-flow TTS practical without relying on a large multi-step sampling budget. Project page: `https://mixispeech.solanai.us/`

## 1 Introduction

Diffusion- and flow-based text-to-speech (TTS) models achieve strong synthesis quality but still require iterative sampling [16,12,9]. Conditional-flow models such as Matcha-TTS [12] already operate with relatively few evaluations, yet pushing inference to the minimum one-step regime remains attractive for further reducing acoustic-generation cost. The key question is therefore whether an already compact multi-step sampler can be compressed to a single evaluation without sacrificing synthesis quality.

⋆ Corresponding author: Cong Tran (congtt@ptit.edu.vn).

This is non-trivial. In our 710-speaker Vietnamese Matcha-TTS model, directly reducing inference from 10 evaluations to one decreases UTMOS22-Strong from 2.2923 to 1.6846 and increases WER from 4.65% to 6.21%. Meanwhile, although Vietnamese TTS has advanced in non-autoregressive, multi-speaker, zero-shot, and large-scale synthesis [1,13,15,25], efficient compression of iterative generative samplers remains underexplored.

We propose MixiSpeech, a progressive state-distillation framework that compresses a conditional-flow acoustic model through a $10 \to 4 \to 2 \to 1$ curriculum. At each stage, the student matches local teacher-integration endpoints using a masked state-matching objective, while retaining the original architecture and multi-speaker conditioning.

On PhoAudiobook [25], the distilled one-step model achieves 2.2678 UTMOS22-Strong, 4.63% WER, and 2.83% CER, close to the 10-step teacher at 2.2923, 4.65%, and 2.72%, while requiring only one acoustic vector-field evaluation.

The main contributions are:

- We show that naive one-step sampling substantially degrades quality in a large-scale 710-speaker Vietnamese conditional-flow TTS setting.
- We propose MixiSpeech, a progressive $10 \to 4 \to 2 \to 1$ state-distillation framework based on local teacher-endpoint matching.
- We demonstrate that the resulting one-step model recovers most of the multi-step teacher quality while improving the quality–computation trade-off.

## 2 Related Work

### 2.1 Efficient and Flow-Based TTS

TTS efficiency has been studied through parallel acoustic generation and simplified synthesis pipelines. FastSpeech 2 [18] enables parallel mel-spectrogram prediction with explicit duration, pitch, and energy modeling, while Glow-TTS [5] uses invertible flows for parallel synthesis with monotonic alignment. VITS [6] further integrates latent-variable modeling, normalizing flows, and adversarial waveform generation in an end-to-end framework. These approaches reduce sequential decoding or pipeline complexity, but do not directly address compression of the sampling trajectory in iterative conditional-flow models.

Diffusion- and flow-based TTS introduce a different efficiency challenge because synthesis requires repeated model evaluations. Grad-TTS [16] generates mel spectrograms through iterative score-based refinement, whereas Matcha-TTS [12] adopts optimal-transport conditional flow matching and an ODE decoder to achieve high-quality synthesis with fewer evaluations. More recent systems such as E2 TTS [3] and F5-TTS [2] further simplify flow-matching TTS and improve training and sampling. In contrast, MixiSpeech focuses on compressing the sampling trajectory of an existing conditional-flow acoustic model into a single vector-field evaluation.

Vietnamese TTS has progressed from non-autoregressive synthesis to multi-speaker and zero-shot generation. Systems from the VLSP 2021 evaluation adopted FastSpeech2-based acoustic models [1,13], while subsequent work explored few-shot multi-speaker adaptation [14] and zero-shot synthesis with diffusion-based acoustic generation [15]. PhoAudiobook further introduced a large-scale Vietnamese speech corpus and evaluated modern zero-shot TTS systems [25]. Unlike these studies, our work targets efficient sampling of a multi-speaker Vietnamese conditional-flow model.

### 2.2 Sampler Distillation

Sampler distillation reduces the number of evaluations required by iterative generative models. DDIM [21] accelerates diffusion sampling through a deterministic non-Markovian process, while progressive distillation [20] repeatedly trains students to replace multiple teacher steps with fewer updates. Consistency models [22] similarly support one-step generation by enforcing consistency along the same trajectory. In TTS, ProDiff [4] distills an $N$-step diffusion teacher into an $N/2$-step student, and SwiftAudio [10] extends one-step distillation to text-to-audio generation.

Building on this line of work, MixiSpeech applies progressive distillation to conditional-flow TTS. Rather than relying on a diffusion-specific parameterization, it trains each student to match endpoints produced by local multi-step teacher integrations, progressively reducing the sampling budget to a single evaluation.

## 3 Method

Figure 1 provides an overview of MixiSpeech. Starting from a conditional-flow acoustic model trained from scratch on the Vietnamese PhoAudiobook training split, our method progressively reduces the global sampling budget through a $10 \rightarrow 4 \rightarrow 2 \rightarrow 1$ distillation curriculum. At each stage, a student is trained to reproduce endpoints returned by local integrations of the preceding frozen teacher using fewer function evaluations, while preserving the same linguistic and speaker conditioning. We first formulate the underlying flow-based sampling process, and then describe the progressive distillation curriculum and its masked state-matching objective.

### 3.1 Problem Formulation

Let $c$ denote the conditioning input supplied to the acoustic generator. In our multi-speaker TTS setting, $c$ collects the text-derived representation, acoustic mask, and speaker conditioning used to guide generation. Let $x_t$ denote the acoustic state at continuous flow time $t \in [0,1]$, where $t=0$ corresponds to the initial noise distribution and $t=1$ corresponds to the generated acoustic sample. In our TTS setting, $x_t$ corresponds to an intermediate mel-spectrogram representation along the generative trajectory, with $x_t \in \mathbb{R}^{F \times L}$, where $F$ is the number of mel-frequency bins and $L$ is the number of acoustic frames.

We draw standard Gaussian noise and apply a fixed sampling temperature:

$$\epsilon \sim \mathcal{N}(0, I), \qquad x_0 = \tau \epsilon, \qquad \tau = 0.15. \tag{1}$$

Thus, $x_0 \sim \mathcal{N}(0, \tau^2 I)$ during distillation and evaluation. The trained-from-scratch base model, parameterized by $\phi$, defines a time-dependent vector field $v_\phi(x_t, t, c)$, which specifies the instantaneous direction and rate of change of the acoustic state at time $t$. The generative trajectory is then governed by the ordinary differential equation

$$\frac{dx_t}{dt} = v_\phi(x_t, t, c), \qquad t \in [0,1], \tag{2}$$

Integrating the learned vector field from $t=0$ to $t=1$ maps the initialized acoustic state toward a sample from the conditional acoustic distribution.

In practice, the continuous trajectory in Eq. (2) is approximated numerically. For example, an $N$-step first-order Euler solver applies

$$\begin{aligned} x_{k+1} &= x_k + \Delta t \, v_\phi(x_k, t_k, c), \\ \Delta t &= \frac{1}{N}, \qquad t_k = \frac{k}{N}. \end{aligned} \tag{3}$$

where $x_k$ denotes the numerical approximation to the continuous state $x_{t_k}$ at time $t_k$, and $N$ is the number of vector-field evaluations in a full uniform-grid sampling run over $[0,1]$.

For an input state $x$ at flow time $a$ and a target time $b \in [a,1]$, let $R = \lceil N(b-a) \rceil$ and $s_r = \min\{a + r/N, b\}$ for $r = 0, \dots, R$. Starting from $z_0 = x$, define

$$\begin{aligned} z_{r+1} = z_r + (s_{r+1} - s_r) v_\phi(z_r, s_r, c), \qquad r = 0, \dots, R-1, \\ \Phi_{\phi,N}^{a \to b}(x, c) = z_R. \end{aligned} \tag{4}$$

Here, $N$ is the nominal Euler resolution and $R$ is the number of evaluations for this local transition. Since student boundaries need not lie on the global $1/N$ teacher grid, each local map uses increments of at most $1/N$ and a shortened final increment. Thus, for non-divisible schedules such as $10 \to 4$, its endpoint need not belong to a single globally discretized $N$-step rollout. For example, $\Phi_{\phi,10}^{0 \to 0.25}(x, c)$ evaluates the teacher at 0, 0.1, and 0.2, with a final increment of 0.05. When $a = b$, the operator is the identity: $R = 0$ and $\Phi_{\phi,N}^{a \to a}(x, c) = x$.

For a full uniform-grid rollout, reducing $N$ lowers inference cost by reducing the number of acoustic vector-field evaluations. A coarser discretization, however, can increase numerical error, and a one- or few-step update need not recover the endpoint of a finer local integration. We therefore train the student's coarse update to match the endpoint returned by a frozen teacher transition, progressively reducing the student's global deployment budget without modifying the underlying acoustic-model architecture.

As illustrated in Figure 1, we use global deployment budgets $10 \to 4 \to 2 \to 1$. The original model first serves as a frozen teacher with nominal local resolution $N = 10$ to generate targets for a 4-step student. The resulting student then supervises a 2-step model, which is finally distilled into a 1-step generator. All stages retain the same text and speaker conditioning, while the state-matching objective used within each stage is introduced below.

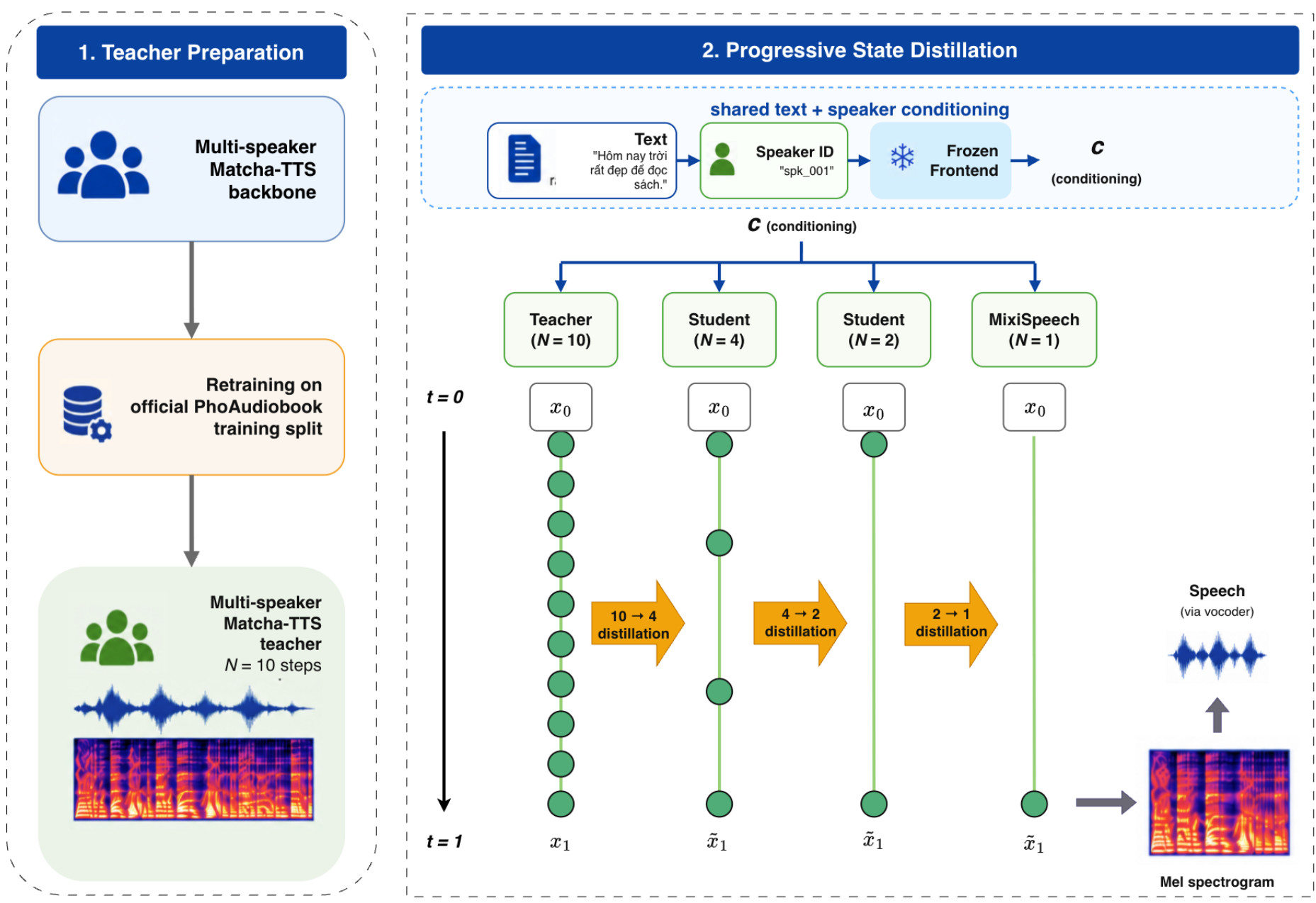

**Fig. 1.** MixiSpeech pipeline: training the multi-speaker Vietnamese teacher from scratch on PhoAudiobook, followed by $10 \to 4 \to 2 \to 1$ global-budget distillation with shared text and speaker conditioning. Intermediate 4- and 2-step students are distilled models, distinct from the direct teacher samplers in Table 1. The final student generates a mel spectrogram for waveform synthesis by the vocoder.

### 3.2 Progressive State Distillation

Directly replacing a multi-step sampler with one update requires a single evaluation to approximate the full sampling interval. Inspired by progressive distillation [20,4], we instead use a decreasing sequence of global sampling budgets

$$N_0 > N_1 > \cdots > N_K = 1. \tag{5}$$

For $k = 1, \dots, K$, stage $k$ distills a teacher with deployment budget $N_{k-1}$ into a student with budget $N_k$, initialized from the teacher. When that student serves as the next teacher, $N_k$ is its nominal map resolution; the prefix and endpoint $R$ counts may differ and sum to more than $N_k$.

Consider a stage with teacher budget and nominal resolution $N$ and student deployment budget $M < N$. Its sampling grid partitions $[0, 1]$ into $M$ intervals with $t_j = j/M$ and $t_{j+1} = (j+1)/M$. We sample $j \sim \text{Uniform}\{0, \dots, M-1\}$ and draw $x_0$ according to Eq. (1). The frozen teacher first produces the local prefix state

$$x_j = \Phi_{\phi,N}^{0 \to t_j}(x_0, c). \tag{6}$$

Starting from the same $x_j$, the teacher supplies the local-integration endpoint target

$$x_{j+1}^T = \Phi_{\phi,N}^{t_j \to t_{j+1}}(x_j, c). \tag{7}$$

Each student interval is sampled and supervised independently. In particular, for non-divisible schedules such as $10 \to 4$, teacher endpoints constructed for different interval indices are not required to compose into a single globally discretized teacher rollout.

The student approximates the complete interval with one update,

$$\hat{x}_{j+1} = x_j + (t_{j+1} - t_j) v_\theta(x_j, t_j, c), \tag{8}$$

At deployment, the $M$-evaluation student is rolled out recursively:

$$\tilde{x}_0 = x_0, \qquad \tilde{x}_{j+1} = \tilde{x}_j + (t_{j+1} - t_j) v_\theta(\tilde{x}_j, t_j, c), \quad j = 0, \dots, M-1. \tag{9}$$

The output is $\tilde{x}_M$; $x_j$ is teacher-forced, whereas $\tilde{x}_j$ is recursive. They may differ at a non-divisible stage even if preceding losses vanish. Thus, this is teacher-forced transition matching, not an exact factorization of a global discrete teacher trajectory. Non-composability occurs only at $10 \to 4$; boundaries align with the teacher grids at $4 \to 2$ and $2 \to 1$.

The training update is optimized by masked endpoint-state matching:

$$\mathcal{L}_{\text{state}} = \frac{\sum_{b,f,\ell} m_{b,\ell} \left( \hat{x}_{j+1,bf\ell} - x_{j+1,bf\ell}^T \right)^2}{F \sum_{b,\ell} m_{b,\ell}}, \qquad F = 80, \tag{10}$$

where $b$, $f$, and $\ell$ index mini-batch samples, mel-frequency bins, and acoustic frames, respectively. The binary mask $m_{b,\ell} \in \{0, 1\}$ marks valid frames and is broadcast across all $F = 80$ mel-frequency bins. Thus, $F \sum_{b,\ell} m_{b,\ell}$ counts the valid acoustic elements, and the loss computes the mean squared error between the student endpoint $\hat{x}_{j+1}$ and local teacher-integration endpoint $x_{j+1}^T$ over non-padded mel-spectrogram elements. No auxiliary recognition, adversarial, perceptual, or speaker-classification loss is used. Since $\Delta t = 1/M$ is fixed within a stage, endpoint-state regression is equivalent up to $\Delta t^2$ to regressing the student field toward $(x_{j+1}^T - x_j)/\Delta t$, the average displacement rate induced by the local teacher integrator.

After each stage, the selected student becomes the frozen teacher for the next stage. Repeating this process increases the interval represented by one evaluation until the final budget reaches one. The deployed student therefore performs

$$\tilde{x}_1 = x_0 + v_\theta(x_0, 0, c), \tag{11}$$

which requires exactly one acoustic vector-field evaluation.

For the multi-speaker instantiation, $c$ also includes the target speaker representation. Each teacher–student training pair uses the same speaker condition, which is retained throughout the trajectory.

The teacher and student start from the same state $x_j$ and conditioning $c$. The teacher target is detached from the computation graph, so the masked state loss in Eq. 10 provides gradients only to the student. For additional implementation details, the pseudocode of the proposed progressive state-distillation algorithm is provided in the supplementary material.

## 4 Experiments

### 4.1 Experimental Setup

**Design.** We compare direct sampling of a single trained-from-scratch multi-speaker Matcha checkpoint with 10/4/2/1 Euler evaluations (MAT-10/4/2/1) against the progressively distilled one-step student (MixiSpeech), under shared conditioning and synthesis settings.

**Data and features.** We use the official PhoAudiobook training/validation/test splits and metadata [25,24]. Training and distillation use only the training split; the validation split is used only for checkpoint selection. For reporting, we sample 100 texts from the official test split and pair each with a randomly chosen speaker from the 710-speaker inventory; the same 100 text–speaker pairs are reused across all systems. Audio is converted to 24 kHz, 80-bin mel spectrograms (FFT/window 1024, hop 256, 0–8000 Hz).

**Backbone and training.** The acoustic model is Matcha-TTS [12] with 20.898M parameters and 710 speakers. Following the standard multi-speaker variant [11], we use a 64-d speaker embedding concatenated into both conditioning paths (encoder: 192+64→256 channels; decoder input: 80 $x_t$+80 $\mu_y$+64→224 channels). The base model is trained from scratch with the native Matcha objectives ($\mathcal{L}_{\text{dur}} + \mathcal{L}_{\text{prior}} + \mathcal{L}_{\text{flow}}$); the lowest-validation-loss checkpoint is used as the frozen teacher.

Progressive distillation runs three stages (10→4, 4→2, 2→1), each for 2,000 updates (batch 6, LR $2 \times 10^{-5}$, weight decay $10^{-4}$), validating every 500 updates (8 validation batches) and selecting the checkpoint with the lowest $\mathcal{L}_{\text{state}}$.

**Synthesis and metrics.** We fix temperature 0.15, length scale 0.96, denoise 0.05, and a shared random seed, and use a frozen 24 kHz HiFi-GAN vocoder [7]. WER/CER are computed with PhoWhisper-large [8] (Whisper-based [17]) after text normalization; UTMOS22-Strong is computed with SpeechMOS v1.2.0 (`utmos22_strong`) [19,23]. We report mean end-to-end inference time and RTF on the same RTX 5090; timing includes frontend, acoustic generation, and vocoding (one run per pair after warm-up, GPU-synchronized). (XTTS-v2 and F5-TTS Vietnamese are reported as contextual external references (details in the supplementary material)).

### 4.2 Main Results

Table 1 reports the automatic quality evaluation on 100 shared test text–speaker pairs. MAT-10, MAT-4, MAT-2, and MAT-1 use the same trained-from-scratch teacher checkpoint with different Euler budgets, while MixiSpeech is the final one-step student obtained through progressive $10 \rightarrow 4 \rightarrow 2 \rightarrow 1$ distillation.

**Table 1.** Automatic quality evaluation. External models are contextual references. NFE denotes acoustic vector-field evaluations. Bold marks the best value within each group; bold with underlining marks the best value across all systems, including ties.

| Model | NFE | UTMOS22-Strong ↑ | WER (%) ↓ | CER (%) ↓ |
|---|---|---|---|---|
| *Multi-step Matcha* | | | | |
| MAT-10 | 10 | 2.2923 | **4.65** | **<u>2.72</u>** |
| MAT-4 | 4 | **2.3206** | 4.77 | 3.18 |
| MAT-2 | 2 | 2.2485 | 5.13 | 3.56 |
| *One-step Matcha* | | | | |
| MAT-1 | 1 | 1.6846 | 6.21 | 3.71 |
| MixiSpeech | 1 | **2.2678** | **<u>4.63</u>** | **2.83** |
| *External references* | | | | |
| F5-TTS Vietnamese (1000h) | 64 | 2.1180 | 32.88 | 23.56 |
| F5-TTS Vietnamese (200h) | 64 | **<u>2.6845</u>** | 35.81 | 17.67 |
| XTTS-v2 Vietnamese | – | 2.2593 | **6.05** | **3.53** |

Directly reducing MAT-10 to one evaluation substantially degrades quality: UTMOS22-Strong decreases from 2.2923 to 1.6846, while WER increases from 4.65% to 6.21%. This result shows that the one-step operating point cannot be obtained reliably by simply using a coarser Euler discretization. In contrast, MixiSpeech achieves 2.2678 UTMOS22-Strong, 4.63% WER, and 2.83% CER with the same one-evaluation budget as MAT-1, substantially recovering the degradation caused by direct one-step sampling.

The comparison with intermediate sampling budgets further illustrates the quality–computation trade-off. Although MixiSpeech uses only one acoustic evaluation, it obtains higher UTMOS22-Strong and lower WER and CER than direct MAT-2 sampling, which requires two evaluations. Relative to MAT-10, MixiSpeech differs by only 0.0245 in UTMOS22-Strong, −0.02 percentage points in WER, and +0.11 percentage points in CER. MAT-4 retains the highest

UTMOS22-Strong among the controlled Matcha systems, while MixiSpeech obtains lower WER and CER with one quarter of its acoustic evaluation budget. Overall, these results indicate that progressive distillation improves the quality–computation trade-off at the one-step operating point rather than simply exchanging synthesis quality for fewer evaluations.

### 4.3 Efficiency Analysis

Table 2 compares the inference efficiency of the evaluated systems. All Matcha-family models retain the same 20.898M acoustic parameters, so the efficiency improvement comes from reducing the number of vector-field evaluations rather than reducing model size.

**Table 2.** Inference efficiency comparison. Inference time is the mean end-to-end synthesis time. Bold marks the best value within each group; bold with underlining marks the best value across all systems, including ties.

| Model | NFE | Params (M) | Infer. (s) ↓ | RTF ↓ |
|---|---|---|---|---|
| *Multi-step Matcha* | | | | |
| MAT-10 | 10 | 20.898 | 0.0662 | 0.0121 |
| MAT-4 | 4 | 20.898 | 0.0433 | 0.0078 |
| MAT-2 | 2 | 20.898 | **0.0356** | **0.0063** |
| *One-step Matcha* | | | | |
| MAT-1 | 1 | 20.898 | **<u>0.0312</u>** | **<u>0.0055</u>** |
| MixiSpeech | 1 | 20.898 | **<u>0.0312</u>** | **<u>0.0055</u>** |
| *External references* | | | | |
| F5-TTS Vietnamese (1000h) | 64 | 335.848 | 1.2678 | 0.2538 |
| F5-TTS Vietnamese (200h) | 64 | 337.107 | **1.1340** | 0.2269 |
| XTTS-v2 Vietnamese | – | 468.643 | 1.2279 | **0.1104** |

### 4.4 Human Evaluation

We conducted a blind listening study with 22 participants, each rating eight samples per system in randomized order with system identities hidden. Using five-point scales, listeners assessed naturalness (1: very unnatural; 5: completely natural), intelligibility (1: unintelligible; 5: completely clear), and speaker similarity (1: not similar; 5: highly similar). Similarity trials included a same-speaker reference recording of a different sentence. All participants completed the study, yielding 176 ratings per system and attribute. We average all listener–utterance ratings; Overall is the unweighted mean of the three attribute means.

MixiSpeech obtains an Overall score of 4.61, versus 4.04 for MAT-1 and 4.66 for MAT-10. It improves on MAT-1 in all attributes and matches MAT-10 in speaker similarity (4.54), but remains below it in naturalness and intelligibility. XTTS-v2 has the highest Overall score (4.68). These descriptive comparisons do not establish statistical significance.

**Table 3.** Blind human evaluation (five-point scales; 176 ratings per system and attribute). Nat., Intell., and Spk. sim. denote naturalness, intelligibility, and speaker similarity, respectively. Overall is their unweighted mean. Higher is better. Bold marks the best mean within each system group; bold with underlining marks the best mean across all eight systems, including ties.

| Model | NFE | Nat. ↑ | Intell. ↑ | Spk. sim. ↑ | Overall ↑ |
|---|---|---|---|---|---|
| *Multi-step Matcha* | | | | | |
| MAT-10 | 10 | <u>**4.74**</u> | **4.69** | **4.54** | **4.66** |
| MAT-4 | 4 | 4.53 | 4.48 | **4.54** | 4.52 |
| MAT-2 | 2 | 4.34 | 4.29 | **4.54** | 4.39 |
| *One-step Matcha* | | | | | |
| MAT-1 | 1 | 3.93 | 3.87 | 4.32 | 4.04 |
| MixiSpeech | 1 | **4.66** | **4.62** | **4.54** | **4.61** |
| *External references* | | | | | |
| F5-TTS Vietnamese (1000h) | 64 | 2.90 | 2.15 | 3.95 | 3.00 |
| F5-TTS Vietnamese (200h) | 64 | 2.60 | 2.05 | 3.70 | 2.78 |
| XTTS-v2 Vietnamese | – | **4.52** | <u>**4.81**</u> | <u>**4.71**</u> | <u>**4.68**</u> |

### 4.5 Qualitative Mel Visualization

Figure 2 shows eight MixiSpeech mel spectrograms with different texts and speaker names. The examples display harmonic bands, temporal transitions, and low-energy intervals. Since the texts differ and no matched teacher or reference spectrograms are shown, these visualizations do not isolate speaker effects or establish speaker fidelity or quality parity.

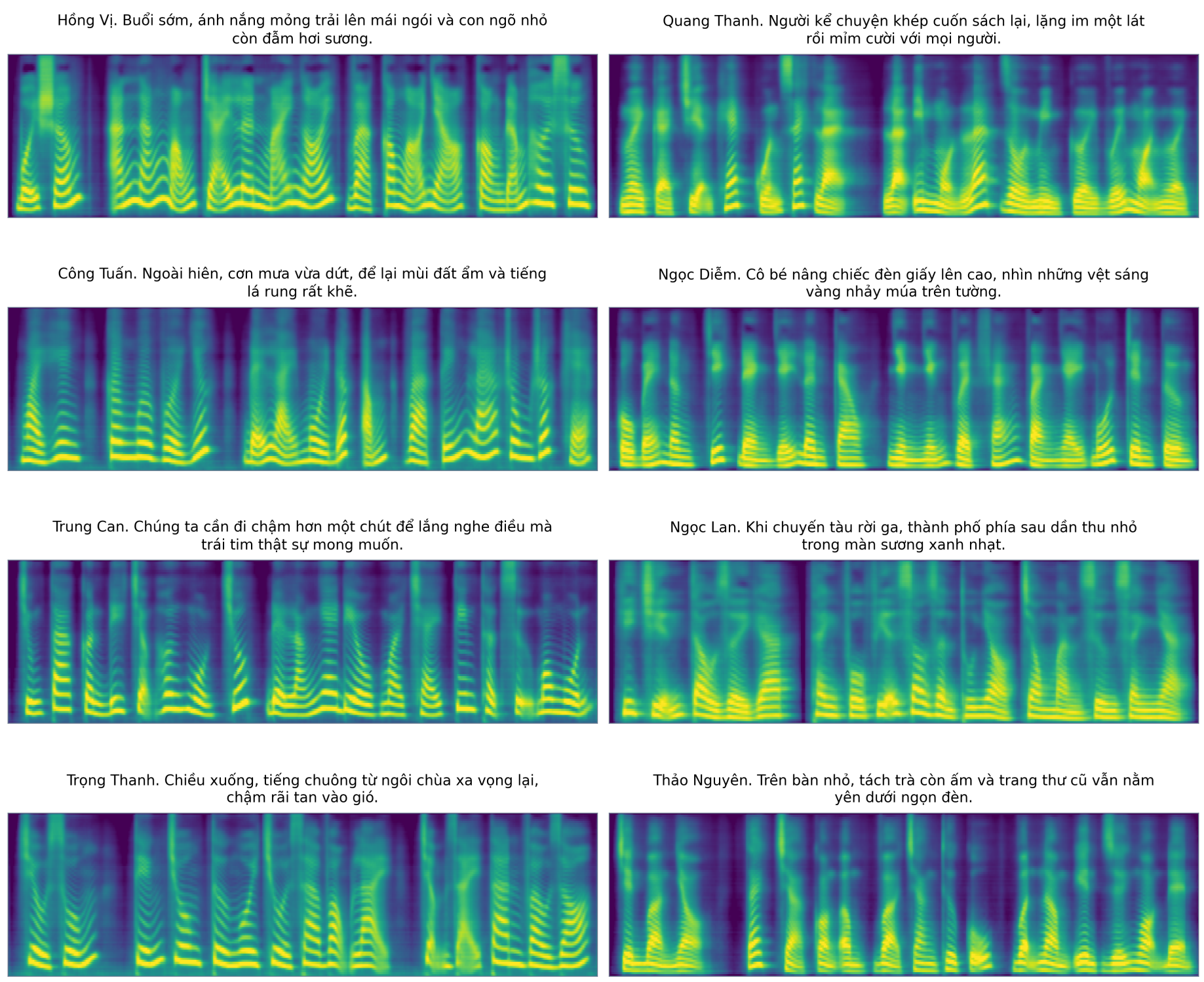

**Fig. 2.** Qualitative MixiSpeech mel examples. Each panel is labeled with a speaker name and Vietnamese text. These examples illustrate generated spectral structure, not a controlled comparison across speakers or systems.

## 5 Conclusion

We introduced MixiSpeech, a progressive state-distillation framework that compresses a multi-step conditional-flow Vietnamese TTS model into a single acoustic vector-field evaluation while preserving most of the teacher's synthesis quality. On 100 shared test text–speaker pairs, the distilled one-step model achieves a UTMOS22-Strong score of 2.2678, a WER of 4.63%, and a CER of 2.83%, closely matching the 10-step teacher while reducing end-to-end inference time from 0.0662 s to 0.0312 s. Human evaluation further shows that MixiSpeech substantially improves upon direct one-step sampling and approaches the perceptual quality of the multi-step teacher.

Beyond Vietnamese multi-speaker synthesis, our results suggest that local endpoint matching can be a simple and effective mechanism for collapsing an ODE-based sampler without altering speaker-conditioning pathway or the backbone architecture. The progressive curriculum also provides practical intermediate operating points (4- and 2-step students) that can be deployed when latency constraints are less strict.

From a systems perspective, reducing the number of acoustic evaluations makes flow-based TTS more attractive for on-device or interactive applications where end-to-end latency and energy consumption are critical. Because distillation preserves the original conditioning interface, the resulting one-step model can be integrated into existing multi-speaker pipelines without additional speaker adaptation or re-training of downstream vocoders.

**Limitations.** The reported quality–efficiency trade-off is established only for 100 PhoAudiobook test texts paired with randomly sampled speakers from the 710-speaker inventory. It does not establish uniform per-speaker performance, generalization to unseen speakers, or robustness to long, code-switched, or atypical inputs. WER, CER, and UTMOS22-Strong are automatic proxies, and the listening study reports mean scores without uncertainty estimates; small differences should therefore not be interpreted as statistically significant. We also evaluate one Matcha-TTS backbone and one $10 \rightarrow 4 \rightarrow 2 \rightarrow 1$ schedule, without a direct $10 \rightarrow 1$ distillation baseline. Per-speaker analyses, uncertainty estimates, direct-schedule ablations, and broader robustness tests would define the scope of these findings more precisely.

In future work, we will study stronger teachers (larger sampling budgets and higher-order solvers), alternative schedules and objectives that better bridge non-divisible step reductions, and broader evaluations covering more speakers, longer-form text, and robustness to challenging inputs (e.g., numerals, abbreviations, and code-switching). We also plan to extend the framework to other conditional-flow and diffusion-based acoustic models, and to validate quality with larger-scale listening tests and statistical analysis.

Overall, MixiSpeech improves the quality–efficiency trade-off at the one-step operating point and makes one-step conditional-flow TTS a practical option for low-latency multi-speaker speech synthesis. We hope this work encourages further research on sampler compression for conditional generative models in under-resourced languages and facilitates broader deployment of multi-speaker Vietnamese TTS in accessibility, education, and interactive voice applications.

# Supplementary Material for MixiSpeech: Progressive State Distillation for One-Step Vietnamese TTS

Tran Quoc Bao Le[1], Binh Mai[1], Hung Dinh[1], Lam Nguyen[1], and Cong Tran[1]

Posts and Telecommunications Institute of Technology, Hanoi, Vietnam

## 1 Vietnamese Teacher Training from Scratch

Before distillation, we train the multi-speaker Matcha base model from scratch [3] on the official PhoAudiobook training split [8,7]. The base model is randomly initialized; no publicly released Matcha-TTS acoustic checkpoint is used to initialize its parameters. For each streamed utterance, the Vietnamese text is phonemized by the same frontend used at synthesis time. The waveform is converted to mono, trimmed, resampled to 24 kHz, and transformed into an 80-bin mel spectrogram. We retain the text, duration, acoustic, and speaker-conditioning paths of the base model. The official validation split is streamed independently and is used only to select the base-model checkpoint.

*Data and preprocessing.* Rows are streamed directly from the official `train` and `validation` splits. We normalize whitespace and phonemize each Vietnamese transcript with the VieNeu frontend. We retain texts with 8–320 characters and waveforms with durations of 2.0–15.5 s. Audio is downmixed to mono, silence-trimmed, resampled to 24 kHz, and converted to 80-bin mel features using FFT size 1024, window length 1024, hop length 256, and frequency range 0–8000 Hz. The speaker map is built from the stream and its embedding is used in both the encoder and decoder conditioning paths. The resulting model covers the same 710-speaker inventory reported in the main paper.

*Optimization and checkpoint selection.* We optimize all teacher parameters using Adam with learning rate $10^{-4}$, zero weight decay, automatic mixed precision, gradient-norm clipping at 1.0, and a shuffle buffer of 10,000 examples. Training uses batch size 8 for up to 130,365 updates. Every 5,000 updates, we compute the mean native Matcha loss on 16 independently streamed validation batches and retain the checkpoint with the lowest validation loss. The test split is never used for optimization or checkpoint selection.

**Algorithm 1** Vietnamese multi-speaker Matcha base-model training from scratch.

**Require:** randomly initialized Matcha-TTS base model $\phi$; official PhoAudiobook splits $\mathcal{D}_{\text{train}}$, $\mathcal{D}_{\text{val}}$; Vietnamese frontend $g$; maximum updates $U$

**Require:** batch size 8; Adam learning rate $10^{-4}$; validation every 5,000 updates on 16 batches; gradient clip 1.0

**Ensure:** selected Vietnamese teacher $\phi_0$

Initialize an empty speaker map $\mathcal{S}$

Initialize Adam optimizer on all parameters of $\phi$

$\mathcal{L}_{\text{best}} \leftarrow \infty$

**for** $u = 1, \ldots, U$ **do**

  Stream a mini-batch of rows $(q, a, s)$ from $\mathcal{D}_{\text{train}}$

  $x \leftarrow g(q)$ ▷ Vietnamese phoneme/token sequence

  $a \leftarrow \text{Resample}_{24\text{kHz}}(\text{Trim}(\text{Mono}(a)))$

  $y \leftarrow \text{Mel}_{80}(a)$; obtain frame mask $m$ and speaker ID $e = \mathcal{S}(s)$

  Discard rows outside the configured text-length or duration ranges

  $(\mathcal{L}_{\text{dur}}, \mathcal{L}_{\text{prior}}, \mathcal{L}_{\text{flow}}, _) \leftarrow \phi(x, y, e, m)$

  $\mathcal{L}_{\text{base}} \leftarrow \mathcal{L}_{\text{dur}} + \mathcal{L}_{\text{prior}} + \mathcal{L}_{\text{flow}}$

  Backpropagate $\mathcal{L}_{\text{base}}$, clip gradients to 1.0, and update $\phi$ with Adam

  **if** $u \bmod 5000 = 0$ **then**

    Stream 16 validation batches from $\mathcal{D}_{\text{val}}$ and compute $\overline{\mathcal{L}}_{\text{val}}$

    **if** $\overline{\mathcal{L}}_{\text{val}} < \mathcal{L}_{\text{best}}$ **then**

      Save $\phi$ as the best base model and set $\mathcal{L}_{\text{best}} \leftarrow \overline{\mathcal{L}}_{\text{val}}$

    **end if**

  **end if**

**end for**

**return** the selected checkpoint as $\phi_0$

The base-model objective in Algorithm 1 is the native Matcha objective used by the training implementation:

$$\mathcal{L}_{\text{base}} = \mathcal{L}_{\text{dur}} + \mathcal{L}_{\text{prior}} + \mathcal{L}_{\text{flow}}. \tag{1}$$

This selected base model is subsequently frozen at the first progressive distillation stage.

## 2 Progressive Distillation Algorithm

Algorithm 2 summarizes the complete $10 \to 4 \to 2 \to 1$ curriculum. For compactness, $\Phi_{\phi,N}^{a \to b}(x, c)$ denotes the frozen teacher's local Euler integration from time $a$ to $b$, using increments no larger than $1/N$ and a shortened final increment when $b$ is not on the nominal teacher grid. The teacher computations are detached, so gradients are applied only to the student parameters.

**Algorithm 2** MixiSpeech progressive state distillation.

**Require:** selected base model $\phi_0$; budgets $(N_0, N_1, N_2, N_3) = (10, 4, 2, 1)$; training data $\mathcal{D}$

**Require:** $U = 2000$ updates per stage; temperature $\tau = 0.15$; validation interval $V = 500$

**for** $k = 1, 2, 3$ **do**
  $N \leftarrow N_{k-1}$; $M \leftarrow N_k$
  Initialize $\theta \leftarrow \phi_{k-1}$ and $\mathcal{L}_{\text{best}} \leftarrow \infty$
  **for** $u = 1, \ldots, U$ **do**
    Sample conditioning inputs $c$ and frame mask $m$ from $\mathcal{D}$
    Sample $\epsilon \sim \mathcal{N}(0, I)$ and set $x_0 \leftarrow \tau\epsilon$
    Sample $j \sim \text{Uniform}\{0, \ldots, M-1\}$
    $t_j \leftarrow j/M$; $t_{j+1} \leftarrow (j+1)/M$
    $x_j \leftarrow \Phi^{0 \to t_j}_{\phi_{k-1},N}(x_0, c)$ ▷ no gradient
    $x^T_{j+1} \leftarrow \Phi^{t_j \to t_{j+1}}_{\phi_{k-1},N}(x_j, c)$ ▷ no gradient
    $\widehat{x}_{j+1} \leftarrow x_j + (t_{j+1} - t_j) v_\theta(x_j, t_j, c)$
    $\mathcal{L}_{\text{state}} \leftarrow \dfrac{\sum_{b,f,\ell} m_{b,\ell} (\widehat{x}_{j+1,bf\ell} - x^T_{j+1,bf\ell})^2}{F \sum_{b,\ell} m_{b,\ell}}$
    Update $\theta$ using $\nabla_\theta \mathcal{L}_{\text{state}}$
    **if** $u \bmod V = 0$ **then**
      Evaluate mean $\mathcal{L}_{\text{state}}$ on eight validation batches
      **if** the validation loss is lower than $\mathcal{L}_{\text{best}}$ **then**
        Save $\theta$ and update $\mathcal{L}_{\text{best}}$
      **end if**
    **end if**
  **end for**
  $\phi_k \leftarrow$ the saved best student; freeze $\phi_k$
**end for**
**return** one-step model $\phi_3$

# 3 Human Listening Evaluation

This section provides the full protocol and aggregation procedure for the human evaluation reported in the main paper. The study compares the five Matcha-family systems and three contextual external systems listed in Table 1.

## 3.1 Evaluated Systems

## 3.2 Listening Protocol

We conducted a blind listening evaluation with 22 participants. Each rated the same eight text–speaker pairs—covering eight distinct target speakers—from every system, matching linguistic and speaker conditions across systems. Presentation order was randomized, system identities were hidden, and all participants completed the study.

The target speakers were Hồng Vị, Quang Thanh, Công Tuấn, Ngọc Diễm, Trung Can, Ngọc Lan, Trọng Thanh, and Thảo Nguyên.

Listeners rated each sample on the three five-point scales summarized in Table 2. For speaker-similarity judgments, the listener was additionally given a ground-truth recording from the same target speaker uttering a different sentence. The reference therefore conveyed speaker identity without repeating the linguistic content of the evaluated sample [7].

**Table 1.** Systems included in the listening evaluation. NFE denotes acoustic vector-field evaluations.

| Group | System | NFE |
|---|---|---|
| Multi-step Matcha | MAT-10 | 10 |
| | MAT-4 | 4 |
| | MAT-2 | 2 |
| One-step Matcha | MAT-1 | 1 |
| | MixiSpeech | 1 |
| External references | F5-TTS Vietnamese (1000h) | 64 |
| | F5-TTS Vietnamese (200h) | 64 |
| | XTTS-v2 Vietnamese | – |

Figure 1 shows the listening interface used in the evaluation.

**Table 2.** Endpoint anchors for the five-point rating scales.

| Attribute | Score 1 | Score 5 |
|---|---|---|
| Naturalness | Very unnatural | Completely natural |
| Intelligibility | Unintelligible | Completely clear |
| Speaker similarity | Not similar | Highly similar |

### 3.3 Score Aggregation

Let $r_{s,a,\ell,u} \in \{1, 2, 3, 4, 5\}$ denote the score assigned to system $s$ and attribute $a$ by listener $\ell$ for evaluated sample $u$. With $L = 22$ listeners and $U = 8$ samples per system, each system receives $LU = 176$ ratings for each attribute. We report the arithmetic mean over all listener–utterance ratings:

$$\bar{r}_{s,a} = \frac{1}{LU} \sum_{\ell=1}^{L} \sum_{u=1}^{U} r_{s,a,\ell,u}. \tag{2}$$

The Overall score is the unweighted mean of the naturalness, intelligibility, and speaker-similarity means:

$$\text{Overall}_s = \frac{1}{3} \left( \bar{r}_{s,\text{nat}} + \bar{r}_{s,\text{intell}} + \bar{r}_{s,\text{spk}} \right). \tag{3}$$

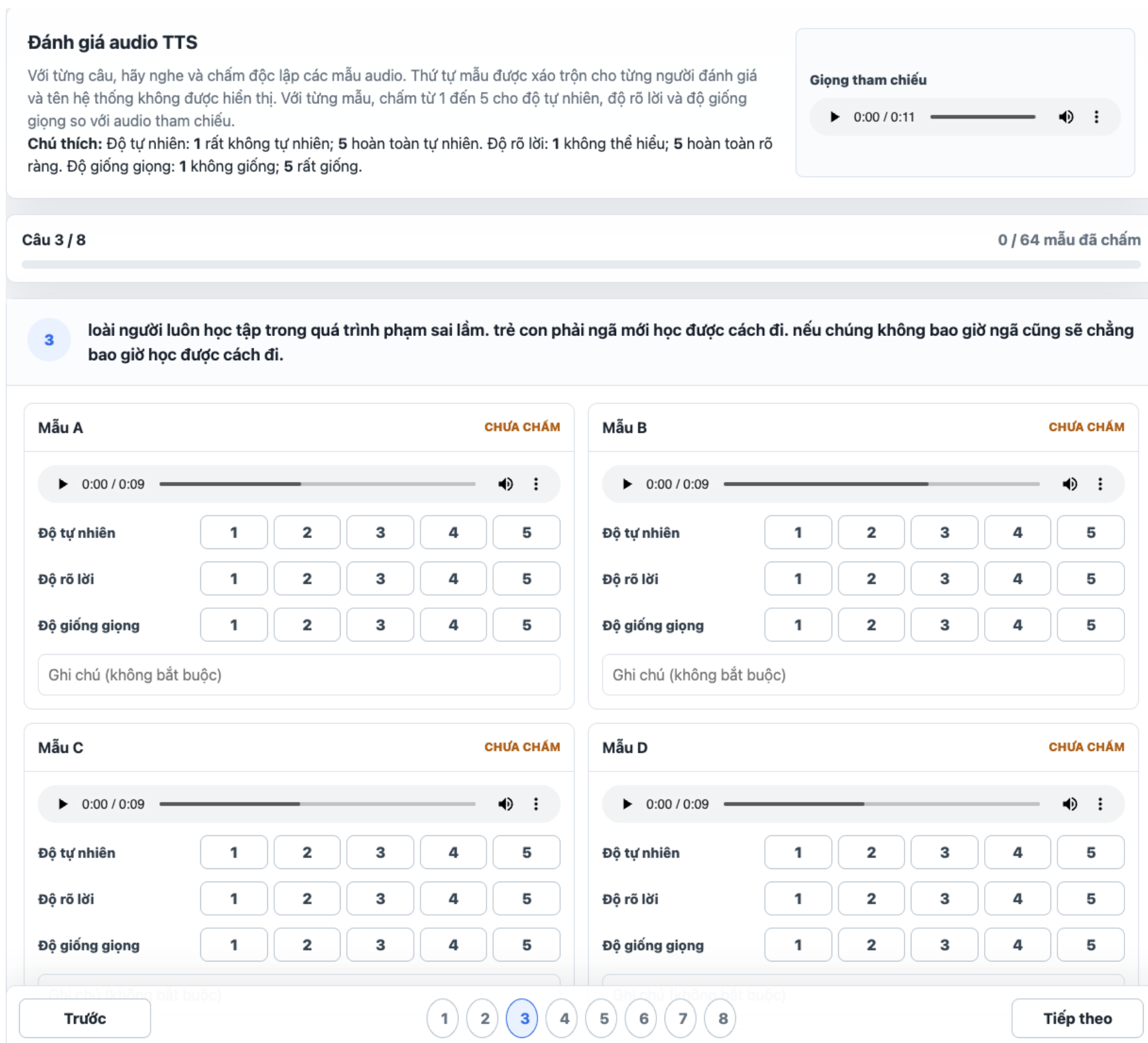

**Fig. 1.** Screenshot of the blind listening-evaluation interface. For each evaluation item, listeners were provided with a ground-truth reference recording and anonymized synthesized samples presented in randomized order. Each sample was rated independently on five-point scales for naturalness, intelligibility, and speaker similarity.

## 4 External Baseline Configuration

The F5-TTS [2] and XTTS-v2 [1] systems are contextual external references rather than architecture-controlled Matcha-family baselines. All prompts were synthesized in Vietnamese. The external systems used ground-truth conditioning recordings for the same eight target speakers listed above. Each reference contained a sentence different from the synthesized evaluation text.

### 4.1 Checkpoints and Implementations

The F5 checkpoints were obtained from the Vietnamese F5-TTS model repository [4], and the XTTS-v2 checkpoint from its Vietnamese model repository [6]. Exact files and architecture configurations are listed in Table 3.

**Table 3.** External-model checkpoints and architecture configurations.

| System | Repository and checkpoint files | Architecture config |
|---|---|---|
| F5-TTS Vietnamese (1000h) | `namprice227/F5-TTS-Vietnamese1000h-Pretrain`: `model_pretrain_1000h.pt` and `pretrain_vocab.txt` | Repository `pretrain_conf.yaml` |
| F5-TTS Vietnamese (200h) | `namprice227/F5-TTS-Vietnamese1000h-Pretrain`: `model_finetune_200h.pt` and `finetune_vocab.txt` | F5-TTS `F5TTS_Base.yaml` |
| XTTS-v2 Vietnamese | `thivux/XTTS-v2-vietnamse`: `best_model.pth`, `config.json`, and `vocab.json` | Repository `config.json` |

The F5 checkpoints were loaded with the project's vendored F5-TTS implementation. Model loading preferred EMA weights and fell back to non-EMA weights only when necessary; both recorded runs loaded successfully with the configurations in Table 3. XTTS-v2 was loaded through the Coqui TTS XTTS implementation with DeepSpeed disabled and inference mode enabled.

### 4.2 F5-TTS Inference

**Table 4.** F5-TTS synthesis settings used for both Vietnamese checkpoints.

| Setting | Value |
|---|---|
| Flow sampling steps (NFE) | 64 |
| Classifier-free guidance strength | 2.0 |
| Sway sampling coefficient | $-1.0$ |
| Speed | 1.0 |
| Cross-fade duration | 0.15 s |
| Language | `vi` |
| Per-prompt random seed | $777 + prompt_index$ |

These settings were passed to F5-TTS `infer_process` with `use_pinyin=False`. The run configuration explicitly records NFE, guidance strength, sway coefficient, and speed; cross-fade duration and the base seed follow the runner defaults. The values are therefore explicit runtime or runner settings, rather than a claim of unmodified checkpoint-native sampling behavior. F5-TTS used its native Vocos vocoder loaded from the selected model configuration [5] and did not use the Matcha 24 kHz HiFi-GAN vocoder.

### 4.3 XTTS-v2 Inference

XTTS-v2 is autoregressive/non-flow in this setup, so NFE is not applicable. The runner called `model.inference` using the checkpoint configuration without manually overriding its sampling values.

**Table 5.** XTTS-v2 Vietnamese synthesis settings.

| Setting | Value |
|---|---:|
| Language | `vi` |
| Temperature | 0.85 |
| Length penalty | 1.0 |
| Repetition penalty | 2.0 |
| Top-$k$ | 50 |
| Top-$p$ | 0.85 |
| GPT conditioning length | 12 |
| Maximum reference length | 10 s |
| Reference normalization | disabled (`sound_norm_refs=false`) |
| Per-prompt random seed | $777 + prompt_index$ |
| Warm-up prompts for timed run | 3 |

Vietnamese text cleaning was patched only to bypass the unavailable Vietnamese cleaner in Coqui TTS while retaining the checkpoint's Vietnamese language token. This was a compatibility modification, not a sampling change. The run metadata does not record whether long-text splitting was enabled, so we make no claim about that option. XTTS-v2 used its native integrated 24 kHz waveform generation pipeline and did not use F5 Vocos or Matcha HiFi-GAN.

### 4.4 Timing and Comparison Scope

All external systems were synthesized on CUDA and timed end-to-end using their native pipelines. No Matcha denoiser or frozen Matcha HiFi-GAN vocoder was applied to their outputs. Because the external systems differ in model scope, conditioning mechanism, and waveform generator, their results serve as contextual references rather than controlled one-step baselines.